\documentclass[slac_one]{revtex4}
\usepackage{graphicx}
\usepackage{fancyhdr}
\begin{document}
\renewcommand{\figurename}{Fig.}
\title{\boldmath
Fourth Generation Leptons and Muon $g-2$}

\author{George Wei-Shu Hou}
\author{Fei-Fan Lee}
\author{Chien-Yi Ma}
\affiliation{
 Department of Physics,
 National Taiwan University,
 Taipei, Taiwan 10617, R.O.C.
}
\begin{abstract}
We consider the contributions to $g_\mu-2$ from fourth generation
neutral and charged leptons, $N$ and $E$, at the one-loop level.
Diagrammatically, there are two types of contributions:
boson-boson-$N$, and $E$-$E$-boson in the loop diagram. In
general, from the Standard Model to the Two-Higgs Doublet Models,
the effect from $N$ is suppressed by off-diagonal lepton mixing
element $V_{N\mu}$. With contribution from $E$, we consider flavor
changing neutral couplings.
\end{abstract}

\maketitle

\thispagestyle{fancy}

\section{MOTIVATION}

It was recently pointed out~\cite{Hou08} that the existence of a
4th generation could have great implications on the baryon
asymmetry of the Universe (BAU). By shifting the Jarlskog
invariant product~\cite{Jarlskog} for $CP$ violation (CPV) of the
3 generation Standard Model (SM3) by one generation, i.e. from
1-2-3 to 2-3-4 quarks, one gains by more than $10^{13}$ in
effective CPV, and may be sufficient for BAU\,! On the other hand,
with renewed interest in the existence of a sequential 4th
generation for CPV studies in B decays (see the references
in~\cite{Hou08}), and with experimental discovery or refutation
expected at the LHC in due time, we turn to the lepton sector.

The difference between the experimental value of muon $g-2$ and the SM3
prediction has been around for some time
now~\cite{Stoeckinger}, i.e.
\begin{eqnarray}
 a_\mu^{\mathrm{exp}}-a_\mu^{\mathrm{SM}}=295(88)\times10^{-11},
 \label{difference}
\end{eqnarray}
where $a_\mu\equiv (g_\mu-2)/2$. The difference is over
3.4$\sigma$, which has aroused a lot of interest. On the other
hand, we have very stringent bounds on lepton flavor violating
(LFV) rare decays, such as~\cite{PDG}
\begin{eqnarray}
 {\cal B}(\mu\to e\gamma) < 1.2 \times 10^{-11},
 \label{mutoegam}
\end{eqnarray}
at 90\% C.L. These limits could be improved further in the near
future. However, the calculations of $a_\mu$ and ${\cal B}(\mu\to
e\gamma)$ are intimately related, coming from similar diagram,
Fig. 1(a), hence giving the similar structures,
\begin{eqnarray}
\epsilon_\lambda q_\nu \sigma^{\lambda\nu}(C_L L+ C_R R).
\end{eqnarray}

\section{EFFECTS FROM NEUTRAL LEPTON $N$ AND CHARGED LEPTON $E$}

\subsection{\boldmath $\text{SM}+N$ }

The contribution from a fourth generation lepton $N$, Fig. 1(b),
has been considered before~\cite{HuoFeng,Lynch1}. We find
\begin{eqnarray}
 a_\mu^\mathrm{SM}(W^{+}W^{-}N)
 \sim233 \times 10^{-11}\,|V_{N\mu}|^2\, F(x),
 \label{WWN}
\end{eqnarray}
where $x = m^2_{N}/M^2_{W}$ and $V_{N\mu}$ is the lepton mixing
matrix element. We depict $F(x)$ versus $x$ in Fig. 1(e). We see that $F(x)$
is a {\it well-behaved} function and {\it bounded}.
From $m_N \gtrsim 90$ GeV~\cite{PDG},
we see that $|V_{N\mu}|$ needs to be
0.7 or higher to reach within $2\sigma$ of Eq.~\ref{difference}.
Considering the stringent constraint from Eq.~\ref{mutoegam},
however, this is clearly unrealistic. We conclude that the
difference of Eq.~\ref{difference} cannot come from the addition
of a 4th neutral lepton $N$.

\subsection{\boldmath $\text{2HDM-II}+N$  }

Going beyond SM, we turn to 2HDM-II (which occurs for MSSM), where
up and down type quarks receive masses from different Higgs
doublets. We find (see Fig. 1(c))
\begin{eqnarray}
 a_\mu^{\rm 2HDM-II}(H^+ H^- N)
 \sim-233 \times 10^{-11}\,|V_{N\mu}|^{2}\,[f_{H^+}(x)+g_{H^+}(x)\cot^2\beta+x_\mu\,q_{H^+}(x)\tan^2\beta],\ \
 \label{HHN}
\end{eqnarray}
where $x=m_{N}^2/M_{{H}^+}^2$, and $x_\mu=m_{\mu}^2/M_{H^+}^2$. We
plot $f_{H^+}(x)$, $g_{H^+}(x)$ and $q_{H^+}(x)$ in Fig.~1(e).
Because $N$ has isospin $+1/2$, large $\cot\beta$ could lead to
enhancement. If we take $|V_{N\mu}\,\cot\beta|^2 \sim 1$ in the
large $\cot\beta$ limit, and if $m_N$ is large compared to
$m_{H^+}$, it could generate a finite, but unfortunately {\it
negative} contribution to $\Delta a_\mu$.

\begin{figure}[t!]
  \includegraphics[width=1.0\textwidth,height=8cm]{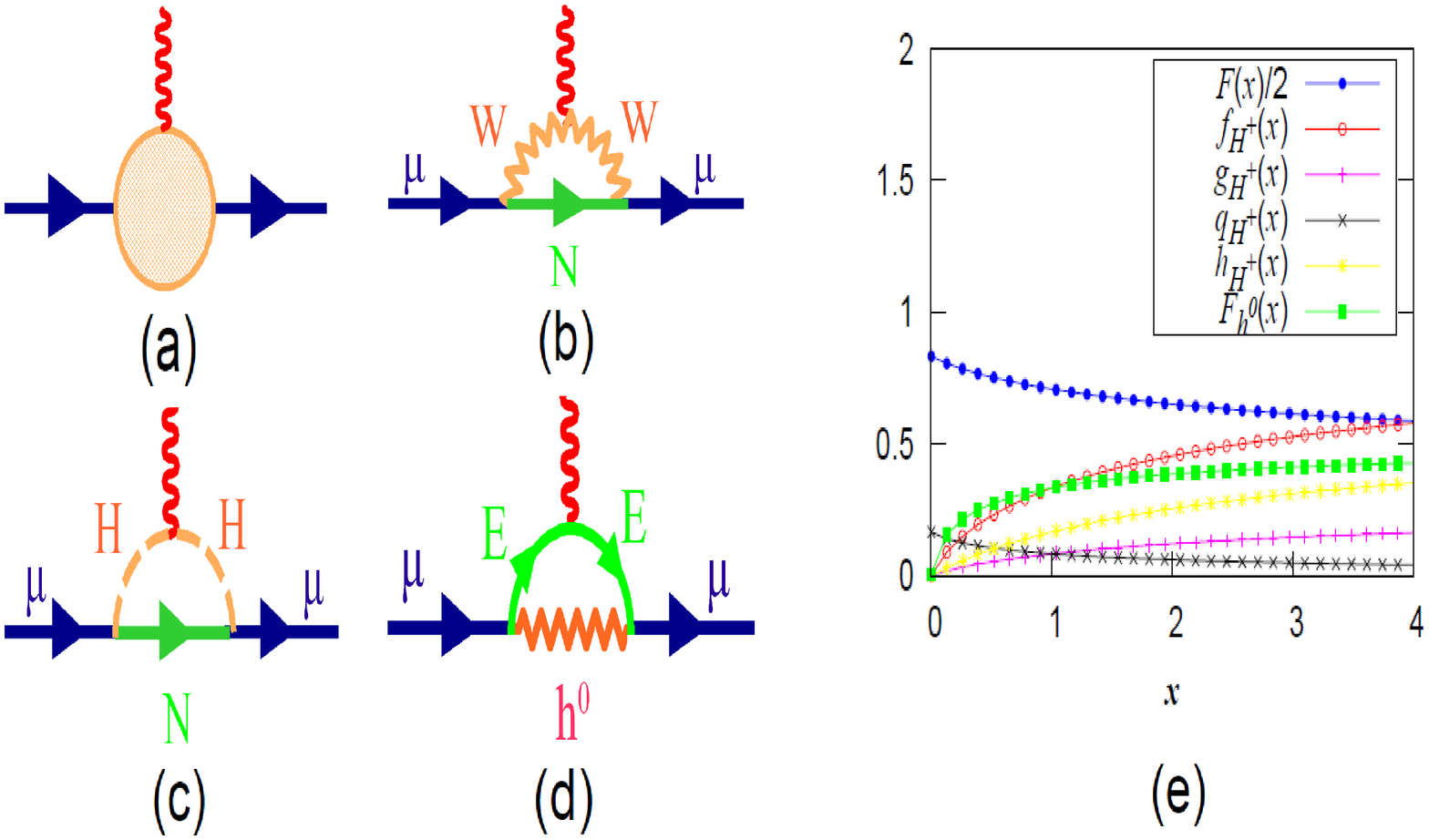}
 \vskip-0.9cm
  \caption{
  (a) loop diagram of $\mu\to e\gamma$ and ${\cal B}(\mu\to e\gamma)$;
  (b) loop diagram from SM+$N$;
  (c) dominant loop diagram from 2HDM-I,II+$N$;
  (d) dominant loop diagram from 2HDM-III+$E$;
  (e) loop functions
  }
\end{figure}

\subsection{\boldmath $\text{2HDM-I}+N$  }

For 2HDM-I, where all quarks receive masses from the same Higgs
doublet, we find
  \begin{eqnarray}
 a_\mu^{\mathrm{2HDM-I}}(H^+H^-N)
 \sim233 \times 10^{-11}\,|V_{N\mu}|^2\tan^2\beta
 [h_{H^+}(x)-x_\mu\,q_{H^+}(x)],\label{HHN2HDMI}
  \end{eqnarray}
 with $x=m^2_{N}/M^2_{H^+}$.
Here we use $v_1(=v\cos\beta)$ to generate all particle masses.
However,  in the 2HDM-I, the $t\bar{t}H^0(h^0)$ coupling relative
to its SM value, $m_t/v$, is given by
$\cos\alpha/\cos\beta\,(\sin\alpha/\cos\beta)$. Large $\tan\beta$
will make the coupling strength $|g_{t\bar{t}H^0}|\gg1$ or
$|g_{t\bar{t}h^0}|\gg1$, and becomes nonperturbative, which
leads us to reject this possibility.

\subsection{\boldmath $\text{2HDM-III}+E$}

In the 2HDM-III, FCNCs are allowed because there exist two
matrices $\eta^{e(\nu)}$ and $\xi^{e(\nu)}$ simultaneously for
each lepton type. To regulate the FCNC in face of stringent
constraints, there is the ansatz suggested by Cheng and
Sher~\cite{ChengSher} for the quark sector, i.e. all
$q_iq_jh^0/H^0/A^0$ couplings have the same form
\begin{eqnarray}
 \Delta_{ij}\frac{\sqrt{m_i m_j}}{v},
 \label{CS}
\end{eqnarray}
where $\Delta_{ij}$ is $\mathcal{O}(1)$. Note that CP-even Higgs
$H^0, h^0$ give {\it positive} contributions to $a_\mu$ but {\it
negative} for $A^0$. Considering the positivity of
Eq.~\ref{difference}, we assume $A^0$ is very heavy and its effect
can be neglected. For sake of illustration, we set $h^0$ to be the
lightest neutral Higgs, and assume no mixing between $H^0$ and
$h^0$, Fig. 1(d). Then we find
\begin{eqnarray}
 \Delta a_\mu^{\mathrm{2HDM-III}} \sim 233 \times 10^{-11}\,
 F_{h^0}(x),\label{EEh}
\end{eqnarray}
where $x=m^2_{{E}}/M^2_{{h}^0}$, and $F_{h^0}(x)$ is given in Fig.
1(e). However, the LFV decay rate in Eq.~\ref{mutoegam} gives a
very stringent constraint. Note that because $a_\mu$ and ${\cal
B}(\mu\to e\gamma)$ come from similar structure of loop diagrams,
their formulas are very closely related. After some organization,
we have
 \begin{eqnarray}
 {\cal B}^{\mathrm{2HDM-III}}(\mu\to e\gamma)
 \sim1.7\times10^{-5} |F_{h^0}(x)|^2\,.\label{meg2HDMIII}
 \end{eqnarray}

Consider the case of $\tau$ in the loop, which is the leading
contribution with 3 generations. Allowing a factor of 2
uncertainty in Eq.~\ref{meg2HDMIII}, we still need
$M_{h^0}>138\,\mathrm{GeV}$ in order to survive
Eq.~\ref{mutoegam}. The MEG experiment will soon push the bound to
$530\,\mathrm{GeV}$. Let us now consider 4 generations. Comparing
Eq.~\ref{difference} with Eq.~\ref{EEh}, we can have
$F_{h^0}(x)\sim1$. However, Eq.~\ref{mutoegam} and
Eq.~\ref{meg2HDMIII} give $F_{h^0}(x)\sim 10^{-3}$, which requires
$m_E\ll M_{h^0}$, which is unlikely. If a 4th generation is found,
the Cheng-Sher ansatz does not seem applicable to the lepton
sector.

\subsection{\boldmath $\text{MSSM}+N+E$  }

Simply put, MSSM doubles the diagrams of SM. The corresponding
loops to $W^+W^-\nu_\mu$ and $\mu\mu Z$ are
$\text{chargino-chargino}$-$\tilde{\nu}_\mu$ and
$\tilde{\mu}$-$\tilde{\mu}$-$\text{neutralino}$ respectively. In
the mass degeneracy limit for superparticles,
$m_{\text{Higgsino}}=m_{\text{Wino}}=M_{\tilde{\nu}_\mu}=M_{\text{SUSY}}$,
and with large $\tan\beta$ (to compensate the extra heaviness of
$M_{\text{SUSY}}$) \cite{Moroi}, we can get a sufficient
contribution to Eq.~\ref{difference}, as has been elucidated in
the literature~\cite{Stoeckinger}.

\section{Summary}

We have discussed some models with 4th generation leptons, and
calculated their impact on $a_\mu$. In the SM, 2HDM-I and II, it
seems that the 4th generation is irrelevant to the $\Delta a_\mu$
puzzle because of the smallness of $|V_{N\nu_\mu}|$. However, this
off-diagonal factor also protects these models from the stringent
${\cal B}(\mu\to e \gamma)$ and ${\cal B}(\tau\to\mu\gamma)$
constraints. For 2HDM-III, there exists a strong conflict with
${\cal B}(\mu\to e\gamma)$ under the Cheng-Sher ansatz with the
4th generation. Hence, if a 4th generation is found, the
Cheng-Sher ansatz cannot hold for the lepton sector. In this
sense, SUSY is favored. Enhancement of $a_\mu$ (diagonal
contribution) and suppression of ${\cal B}(\mu\to e\gamma)$
(off-diagonal contribution) in the MSSM are both similar to the
SM. Since large $\tan\beta$ suppresses the {\it negative}
contribution from $H^+H^-N$, MSSM and 4th generation can coexist.

\begin{acknowledgments}
This work is supported in part by the National Science Council of
R.O.C. under grant numbers NSC96-2739-M-002-005,
NSC96-2811-M-002-042, and NSC96-2811-M-002-097.
\end{acknowledgments}

\end{document}